\documentclass[%
 reprint,
 amsmath,amssymb,
 aps,
prb,
floatfix,showpacs
]{revtex4-1}

\usepackage{graphicx}
\usepackage{dcolumn}
\usepackage{bm}
\usepackage{color}
\usepackage{gensymb}
\usepackage{array}
\newcolumntype{P}[1]{>{\centering\arraybackslash}p{#1}}

\usepackage[normalem]{ulem}

\begin{document}

\preprint{APS/123-QED}

\title{Experimental observation of hybrid TE-TM polarized surface waves supported by hyperbolic metasurface}

\author{Oleh Y. Yermakov}%
\email{o.yermakov@metalab.ifmo.ru}
\author{Anna A. Hurshkainen}%
\author{Dmitry~A.~Dobrykh}
\author{Polina~V.~Kapitanova}%
\author{Ivan~V.~Iorsh}%
\author{Stanislav~B.~Glybovski}%
\email{s.glybovski@metalab.ifmo.ru}
\author{Andrey A. Bogdanov}%
\affiliation{Department of Nanophotonics and Metamaterials, ITMO University, St. Petersburg 197101, Russia}%




\date{\today}

\begin{abstract}
Hyperbolic metasurfaces have gained significant attention due to their extraordinary electromagnetic properties to control propagating plane waves, but the excitation and propagation of the surface plasmon-polaritons at hyperbolic metasurfaces, called \textit{hyperbolic plasmons}, have been experimentally observed just recently. However, the advantages of the hyperbolic plasmons, such as \textit{hybrid TE-TM polarization} discussed below, are not yet fully revealed and analyzed. 
In this paper we focus on the numerical and experimental characterization of surface waves in the frequency range from 2 to 8 GHz supported by a hyperbolic metasurface composed of anisotropic metallic Jerusalem crosses printed on a thin dielectric substrate. We show different shapes of equal frequency contours, which correspond to a plethora of excitation and propagation regimes of surface waves. The principal novelty of this work consists in the experimental demonstration of the surface waves with a hybrid, i.e. mixed TE-TM, polarization. Surface waves with a hybrid polarization are the promising tool in a number of applications and phenomena including polarization converters, plasmonic sensors, plasmon steering over a surface, optical forces, spin-orbit photonics, and highlight the impact for the on-chip and planar networks.

\end{abstract}

\maketitle


\section{Introduction}

The manipulation and control of electromagnetic waves is one of the central aim of photonics applications. Impressive results in this area have been achieved with metasurfaces (2D metamaterials) -- periodic planar arrays of resonant subwavelength building blocks (meta-atoms). By appropriately engineering the shape and the spatial arrangement of these building blocks, it is possible to design a metasurface with specified properties. Metasurfaces are widely used to manage the polarization, transmission and reflection of electromagnetic waves~\cite{zhao2011manipulating,pfeiffer2013metamaterial,pors2013plasmonic,moitra2014experimental}. Moreover, metasurfaces can serve as optical control devices such as frequency selectors, perfect absorbers, antennas, polarization transformers, switchers, sensors, etc~\cite{holloway2012overview,glybovski2016}. It has been pointed out that metasurfaces provide an unrivalled control over the dispersion and polarization of surface plasmon-polaritons (SPPs)~\cite{yu2014flat,mencagli2015surface,gomez2015hyperbolic,yermakov2015hybrid,takayama2017photonic}. At the same time metasurfaces are fully compatible with photonic devices and integrated optical circuits due to their relative simplicity, planar geometry, light weight and low losses in comparison to 3D metamaterials~\cite{kildishev2013planar,yu2014flat}.    

The studies of hyperbolic metamaterials have attracted a great interest due to their unique electromagnetic properties such as negative refraction, canalization and large density of states~\cite{cortes2012quantum,poddubny2013hyperbolic}. They find the corresponding applications, for instance, hyperlens~\cite{fang2005sub,jacob2006optical}, enhanced spontaneous emission~\cite{poddubny2011spontaneous,sreekanth2013directional,lu2014enhancing}, perfect absorbers~\cite{tumkur2012control,nefedov2013total}, routing of optical signals~\cite{kapitanova2014photonic}. 

Two-dimensional structures with hyperbolic response also provide unusual properties in optical, THz and radiofrequency ranges. A.~A.~High et al. have demonstrated that a hyperbolic metasurface based on a silver/air grating provides the negative refraction and diffraction-free propagation of the SPP as well as the plasmonic spin Hall effect in the visible range~\cite{high2015visible}. Another appealing feature is that the spectrum of a hyperbolic metasurface supports two mixed TE-TM polarized modes, which are usually called the \textit{hyperbolic plasmons}~\cite{gomez2015hyperbolic,yermakov2015hybrid}. This \textit{hybridization} of TE and TM polarizations for surface waves is the consequence of the anisotropy. It should be mentioned that at a microscopic scale comparable with the size of the unit cell, i.e. in the local near-field, the structure of electromagnetic field is always hybrid and complicated~\cite{burresi2009observation,de2014optical}. However, at the wavelengths that are much larger than the period of the structure, the electromagnetic field over the unit cell is averaged. So, in the case of a hyperbolic metasurface we deal with a hybrid polarization of surface waves in the macroscopic point of view. In this sense, metasurface can be described by the averaged effective quantity, for instance, conductivity~\cite{yermakov2018effective}.

The keynote advantage of the hyperbolic plasmons with hybrid TE-TM polarization over the conventional TM and TE ones is the full control over propagation direction, polarization and wavefront shape. Hybrid nature of hyperbolic SPPs makes possible to engineer their polarization structure providing efficient interaction with both electric and magnetic resonant particles. The dispersion of the surface waves at hyperbolic metasurface depends on the propagation direction similarly to the Dyakonov surface states~\cite{dyakonov1988new,takayama2017midinfrared,takayama2018experimental}. Highly directional and canalization propagation regimes can be used for the routing of SPPs, in-plane beam steering, subwavelength imaging and hyperlenses~\cite{sedighy2013wideband,high2015visible,gomez2016flatland,correas2017plasmon}. Polarization degree of freedom inherent to hyperbolic plasmons could be used for the enhancement of light-mater interaction with polarization-sensitive quasi-particle such as exciton-polariton. Polarization and optical spin angular momentum of the surface waves localized at hyperbolic metasurfaces can be continuously changed on demand~\cite{yermakov2016spin}. Besides, the complicated fields distribution of hyperbolic plasmons can be constructed in order to manipulate force and torque acting on a particle in the vicinity of a surface~\cite{rodriguez2015lateral, yermakov2016spin}. Furthermore, polarization-dependent surface plasmons are the promising platform for the sensing, particles sorting, enhanced chiral spectroscopy and quantum information science~\cite{sreekanth2016extreme,jiang2017multifunctional,hu2018metasurface,shkondin2018high}. Moreover, the interplay between hyperbolic dispersion and hybrid character of the polarization leads to the rich variation of the equal frequency contours (EFCs), which results in the multiplicity of the excitation and propagation surface waves regimes~\cite{yermakov2015hybrid,gomez2015hyperbolic,mencagli2015surface,gomez2016flatland,nemilentsau2016anisotropic,samusev2017polarization}.

The negative mode index of TM surface mode has been shown experimentally with a mushroom-type impedance surface in the microwave range~\cite{dockrey2016direct}. The spoof plasmons with hyperbolic dispersion excited by an electric dipole and propagating along an ultra-anisotropic metasurface composed of H-shaped slit inclusions have been recently detected in microwaves~\cite{yang2017hyperbolic}. These hyperbolic spoof SPPs exhibit a sharp transition from elliptical to hyperbolic equal frequency contours (EFCs), which can be widely used in planar photonic devices. Furthermore, we demonstrated a short time ago the propagation at optical frequencies of TE and TM plasmon modes supported by the anisotropic metasurface composed of the resonant plasmonic elliptical nanodisks~\cite{samusev2017polarization}. These polarization-dependent hyperbolic plasmons possess highly nontrivial and rich set of EFCs. 

In this work, we focus on the numerical and experimental characterization of the polarization properties of surface waves supported by a hyperbolic metasurface in the microwave frequency range (from 2 to 8 GHz), especially on their hybrid behaviour. The metasurface is represented by a two-dimensional array of anisotropic copper Jerusalem crosses printed on a thin dielectric substrate. We observe the propagation of two localized surface modes, which have mixed TE-TM polarization (quasi-TE or quasi-TM) due to the anisotropy of the metasurface. It should be highlighted that we explicitly demonstrate the theoretically predicted hybridization of surface modes, which has not been experimentally confirmed to the best of our knowledge. The evolution of equal frequency contours, corresponding to the different SPP propagation regimes, for each eigenmode is retrieved and investigated. Finally, drastical changes of the EFC shapes, called \textit{topological transitions}, are found and confirmed experimentally.

\section{Design and Experimental Study}

\begin{figure*}[bt]
\centering
  \includegraphics[width=0.98\textwidth]{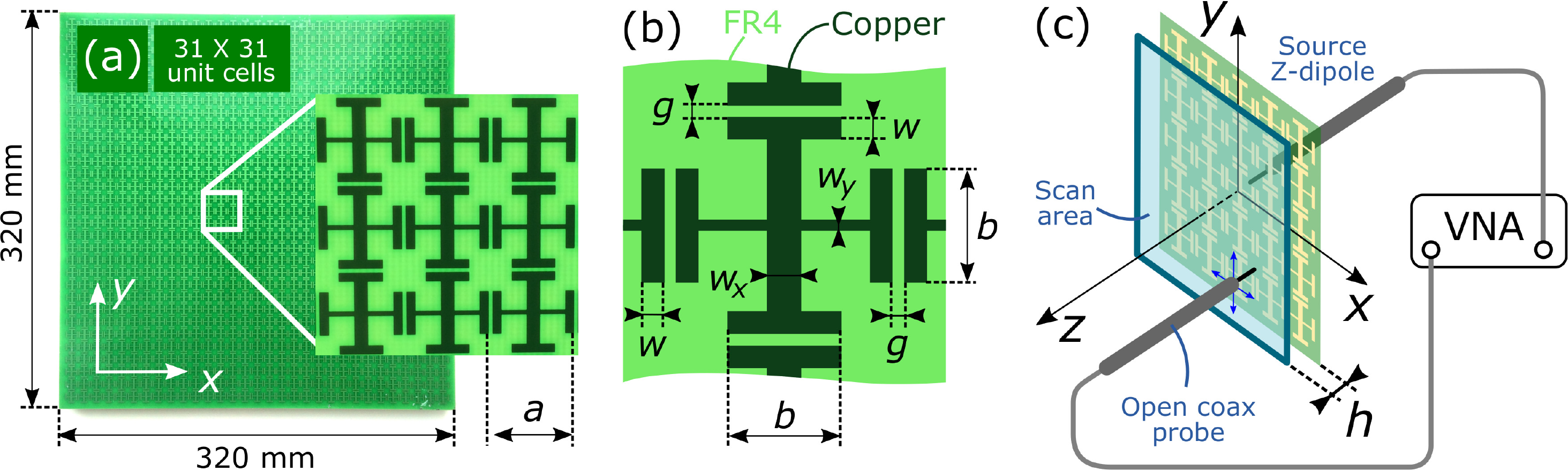}
  \caption{(a) Photograph of the metasurface prototype. The inset demonstrates the photograph of the selected region with 3 by 3 unit cells. (b) The geometry of the metasurface unit cell. (c) Schematic view of the experimental setup for the configuration "probe-probe": vertical electric dipole is used as an excitation source and a normally oriented coax probe is used to detect the vertical component of the electric field. The studied metasurface is located in the XY-plane.} 
 \label{fig:experiment}
\end{figure*}

The metasurface prototype has been fabricated using the printed circuit board technology.  The photograph of the metasurface, which can be classified as an anisotropic frequency-selective surface, is shown in Fig.~\ref{fig:experiment}a. It consists of 31$\times$31 unit cells arranged with the periodicity of $a=10$ mm and has the overall dimensions of 320$\times$320 mm$^2$. The unit cells in a shape of an anisotropic Jerusalem cross  (Fig.~\ref{fig:experiment}b) were etched into the copper layer of 1.5-mm thick FR4 substrate ($\varepsilon_r=4.3$, $\tan\delta=0.02$). 
The central vertical and horizontal copper traces of each cross have different widths of $w_x=1.5$~mm and $w_y=0.5$~mm, correspondingly, which govern the anisotropic grid impedance properties. The series resonances of the metasurface occur at 5 GHz and 6 GHz for the X-polarized and Y-polarized impinging plane wave, respectively. It is explained by higher unit-cell inductance for the currents flowing along the horizontal central traces, that are thinner than the vertical ones. All side traces of crosses have the same width of $w=1$~mm. All the capacitive gaps have the same width of $g=0.5$~mm. All four side traces of the cross are of the same length $b=5$~mm. For near-field experimental study the metasurface was located in an anechoic chamber and tested using a near-field scanner.

It is well-known that electric and magnetic fields of TM and TE surface waves due to their symmetry can be excited and measured by the electric and magnetic dipoles, respectively, placed normally with respect to the metasurface. In our experiment, we measure the same normal component of the field vectors. Due to the hybrid nature of the surface waves, both the electric and magnetic normal components have to be measured. In the experiment the electric dipole can be represented by a vertical coaxial probe, while the magnetic one - by a horizontal loop (metasurface is assumed to be in a horizontal plane) as it was done in Refs.~[\onlinecite{dockrey2016direct,yang2017hyperbolic}].

First, we measured the field patterns of quasi-TM surface waves excited at the metasurface by a vertical electric dipole. In this case, the normal electric field component still dominates and it is crucial to be detected. At the same time, in contrast to purely TM waves, a non-zero normal magnetic field component exists as well and should be measured as well. The schematic view of the experimental setup, used for measuring the normal electric field component of quasi-TM surface waves, is depicted in Fig.~\ref{fig:experiment}c. An open end of a coaxial cable with elongated core with the length of 0.5~mm was used as a vertical electric dipole probe. It was fixed at the 20~mm distance nearby the geometric metasurface center and connected to the first port of Agilent E8362C Vector Network Analizer (VNA). The identical open end coaxial cable, connected to the second port of the VNA, was used as a receiving electric dipole (probe). The field was measured by moving the probe parallel to the metasurface (XY-plane) over the scan area of 320$\times$320 mm$^2$ with 3~mm step at the distance of $h=20$~mm. At each point the complex value of the transmission coefficient between the source and probe was measured resulting in the mapping of the metasurface near-field. Similarly, the dominant normal magnetic field of quasi-TE surface modes, launched by a vertical magnetic dipole source, was measured with a magnetic dipole. To observe the hybrid nature of the surface waves, the normal electric field component was measured for the same magnetic dipole source as well using the above mentioned coaxial probe. We have used two identical Faraday loops~\cite{jackson2007classical,pendry1999magnetism} with the diameter of 6~mm as the magnetic dipole source and probe. The plane of the loop was parallel to the plane of the metasurface and was located 10~mm away from it.

In the measurements of the fabricated metasurface we separately consider the excitations by a probe (vertical electric dipole) and a loop (vertical magnetic dipole) sources. In each case we detect both the signals of a probe and a loop receiver, i.e. the normal field components $E_z$ and $H_z$, according to Table~1. The measured field patterns are compared to the full-wave numerical simulations of a whole metasurface ($31\times 31$ unit cells) performed with time-domain solver of CST Microwave Studio. One of the four configurations, i.e. probe-probe is depicted schematically in Fig.~1c. To retrieve the EFCs related to different setup configurations the 2D Fast Fourier Transform has been applied to the measured complex field patterns, following the direct observation approach from the work~[\onlinecite{dockrey2016direct}].

\begin{table}[htbp]
\centering
\begin{tabular}{P{2.3cm} P{3.4cm} P{2.2cm}}
\hline \hline
\textbf{Configuration name} & \textbf{Excitation source} & \textbf{Detected surface mode} \\
\hline 
loop-loop & vertical magnetic dipole & quasi-TE \\
loop-probe & vertical electric dipole & quasi-TE \\
probe-loop & vertical magnetic dipole & quasi-TM \\
probe-probe & vertical electric dipole & quasi-TM \\
\hline \hline
\end{tabular}
\caption{Four experimental excitation and probing configurations.}
\end{table}

\section{Results and Discussions}

\begin{figure}[b]
  \centering
  \includegraphics[width=0.98\linewidth]{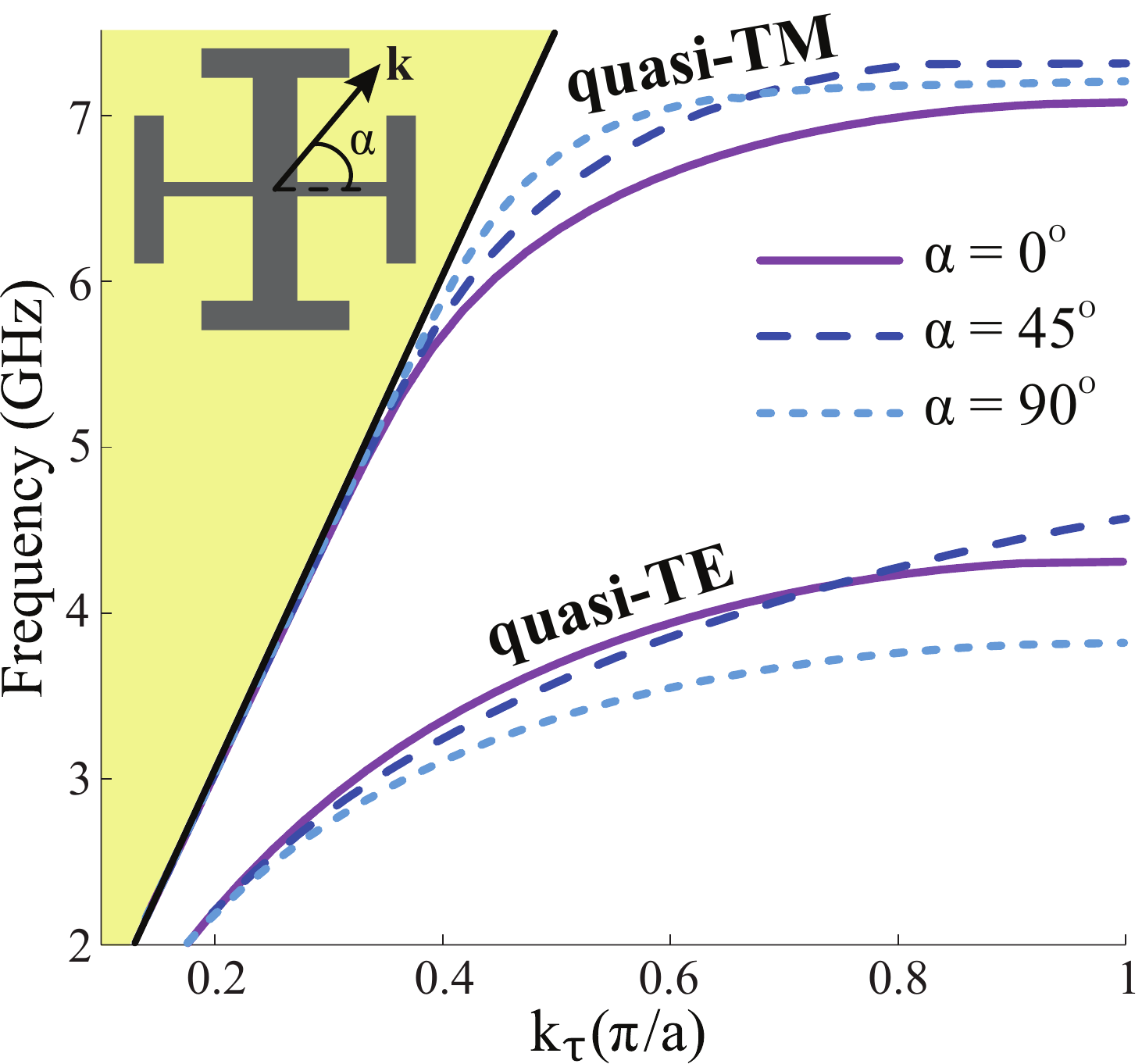}
  \caption{Numerically calculated dependence of the in-plane wavevector $k_{\tau}$ on frequency for the surface waves localized at metasurface, depicted in Fig.~1.}
  \label{fig:dispersion_cst}
\end{figure}

Usually, hyperbolic metasurfaces are described in terms of two-dimensional effective surface (or grid) impedance~\cite{mencagli2015surface} or electric conductivity~\cite{yermakov2015hybrid,gomez2015hyperbolic} in the form of an anisotropic diagonal tensor. However, the metasurface structure under consideration cannot be described within the local effective medium approximation (homogenization procedure) due to the strong non-dipolar and non-local interaction between capacitive edges of Jerusalem crosses in neighboring unit cells. Nevertheless, the effective surface conductivity can well characterize this structure at low frequencies below the bandgap~\cite{yang2017hyperbolic}. We calculate the dispersion of surface waves localized at the proposed metasurface design taking into account non-local effects in CST Microwave Studio by using Eigenmode Solver. The numerically calculated dispersion laws for different propagation directions $\alpha$ are shown in Fig.~\ref{fig:dispersion_cst}. One can see that the surface plasmon resonance (SPR) frequency increases from $\alpha = 0^\circ$ to $\alpha = 45^\circ$ and, then, decreases from $\alpha = 45^\circ$ to $\alpha = 90^\circ$, which is a consequence of the cross geometry. According to the effective medium approximation the SPR for any propagation angles for the first mode should be observed between 5 and 6~GHz, i.e. the frequencies at which the peaks of reflection and transmission coefficients under normal incidence for X- and Y-polarization take place~\cite{yermakov2015hybrid}. The numerical calculations show that SPR for the first surface mode (quasi-TE) is below 5~GHz (between 3.8 and 4.6~GHz), which confirms the shift related to the non-dipolar and non-local contributions. According to the theoretical analysis performed in the Ref.~[\onlinecite{yermakov2015hybrid}] for a metasurface with a similar electromagnetic response under normal incidence, both eigenmodes supported by an anisotropic hyperbolic metasurface have hybrid TE-TM polarization, but we can distinguish the surface modes as \textit{quasi-TE} and \textit{quasi-TM}.

\begin{figure}[t]
  \centering
  \includegraphics[width=0.92\linewidth]{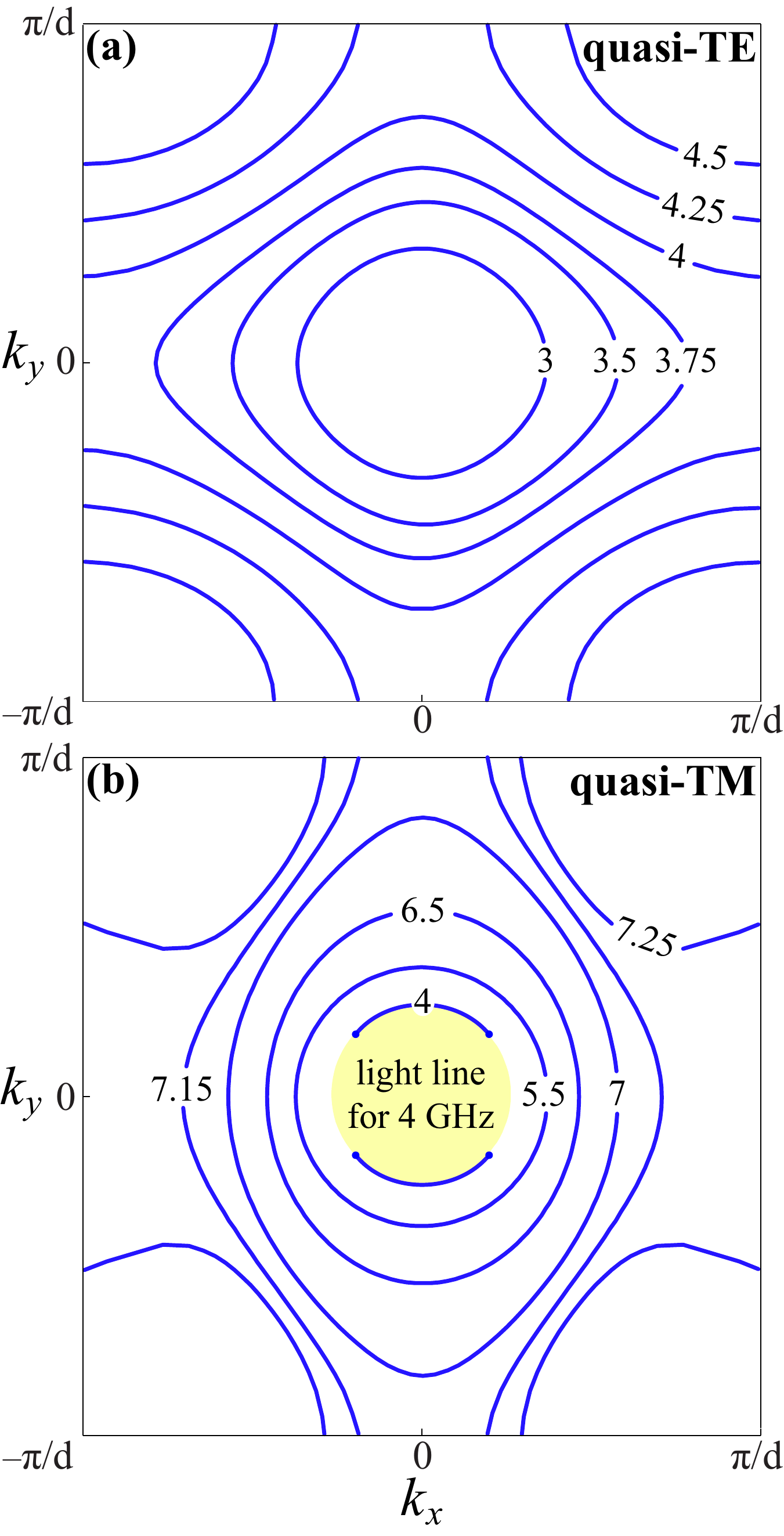}
  \caption{The equal frequency contours for the surface waves excited at the metasurface, depicted in Fig.~1, calculated numerically in CST Microwave Studio.}
  \label{iso_cst}
\end{figure}

Surface modes properties of the considered anisotropic metasurface are manifested most clearly in the simulated EFCs plotted within the first Brillouin zone (Fig.~\ref{iso_cst}). At low frequencies quasi-TE mode dispersion is close to the light line and the corresponding EFC is very similar to a circle (3~GHz). With the increase of the frequency the circle transforms into the ellipse (3.5 and 3.75~GHz) elongated along $x$-direction. Between 3.75 and 4~GHz, the EFC of the quasi-TE mode becomes open, which leads to a forbidden range of propagation directions. Such a pivotal change of EFC is called \textit{topological transition}~\cite{krishnamoorthy2012topological}. At higher frequencies (4.25 and 4.5~GHz) the propagation of TE-like surface mode is possible only in a narrow angular range in the vicinity of the diagonals of the Brillouin zone, that means the high density of optical states (like in the case of a flat band).  This results in pronounced directivity of surface waves, which can lead to the significant emission enhancement from emitters disposed in the vicinity of a hyperbolic metasurface~\cite{kildishev2013planar}.

Similar evolution of EFCs at higher frequencies takes place for quasi-TM mode (Fig.~\ref{iso_cst}b). The main difference is in the elliptical EFCs orientation. Importantly, quasi-TM mode dispersion starts from the light line, i.e. has a cut-off point, in contrast to the quasi-TE one~\cite{yermakov2015hybrid}, which is shown at the frequency $f = 4$~GHz. In this case the EFC can be classified as an arc, whereas the intersection point of the arc and light circle corresponds to the cut-off point. With the increase of the frequency arcs transform to the ellipses (5.5, 6.5 and 7~GHz) elongated along $y$-direction. Between 7 and 7.15~GHz, angular bandgap of the quasi-TM mode emerges. At higher frequencies the range of the forbidden directions increases, which is a consequence of the EFCs in the form of near-corner arcs (7.25~GHz).

\begin{figure}[htb]
  \centering
  \includegraphics[width=0.92\linewidth]{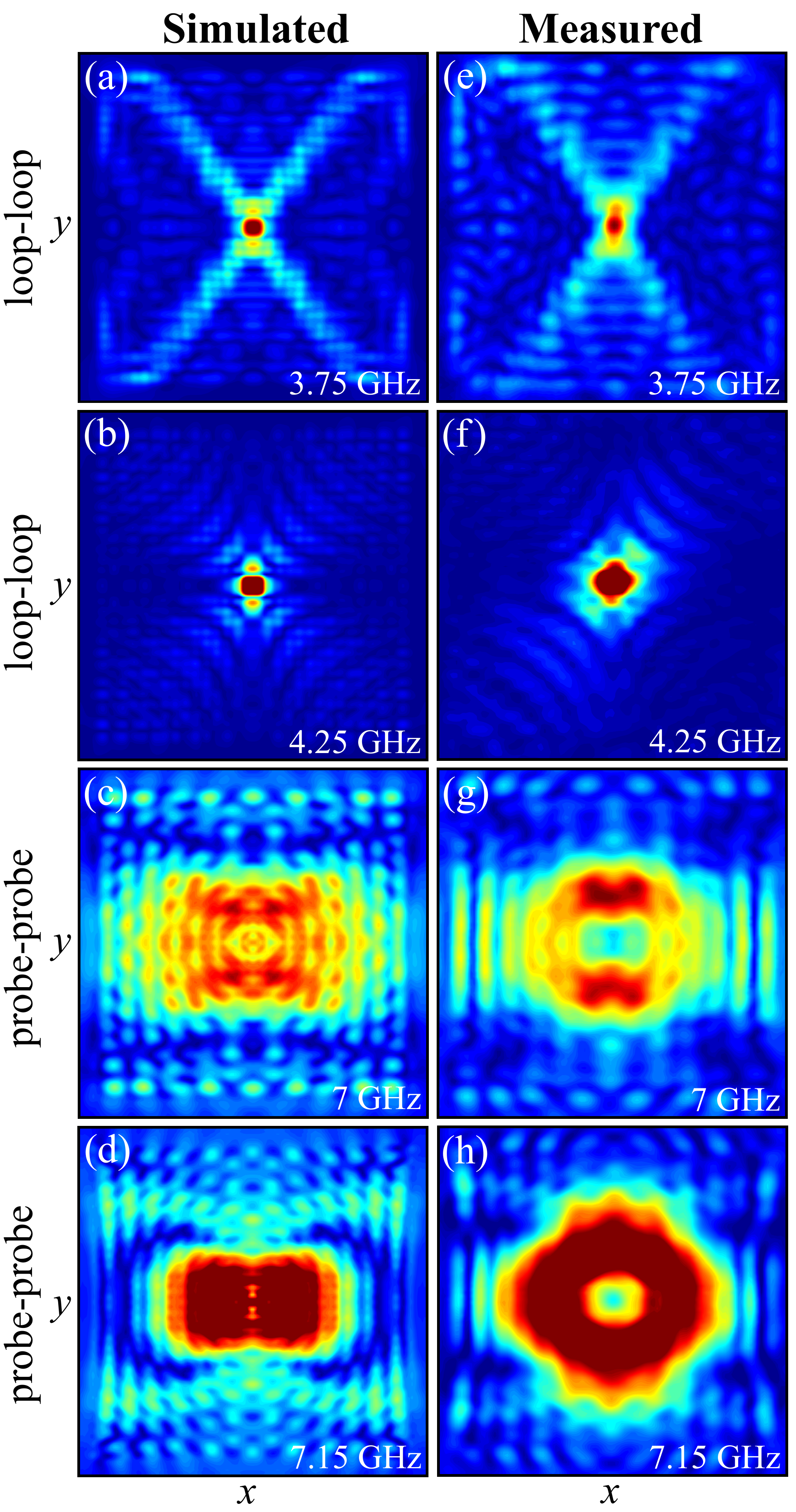}
  \caption{The numerical (a-d) and experimental (e-h) distributions of the perpendicular component of magnetic (a,b,e,f) and electric (c,d,g,h) fields for different frequencies.}
  \label{fields}
\end{figure}

The experimental and numerical distributions of the perpendicular component of magnetic and electric fields with loop-loop and probe-probe setup configurations, respectively, are presented in Fig.~\ref{fields}. The shapes of wavefronts shown in Fig.~\ref{fields} can be explained by using the equal frequency contours shown in Fig.~\ref{iso_cst}. Figures~\ref{fields}a and \ref{fields}e correspond to the sharply elongated (along $x$-direction) ellipse close to the rhombic form in $k$-space (Fig.~\ref{iso_cst}a). In real space, the rhombic EFC corresponds to the dipole radiation in four distinct angles with almost flat wavefronts~\cite{yermakov2015hybrid}. The surface wave propagation shown in Figs.~\ref{fields}b and \ref{fields}f is four-directional, which corresponds to the near-corner arcs in EFCs (Fig.~\ref{iso_cst}a). The electric field distribution of quasi-TM mode shown in Figs.~\ref{fields}c and \ref{fields}g can be classified as one-directional (along $x$-axis) propagation. The corresponding EFC in reciprocal space represents the elongated (along $y$-direction) ellipse tending to the discontinuity (Fig.~\ref{iso_cst}b). At 7.15~GHz the EFC becomes open (Fig.~\ref{iso_cst}b), which determinates the topological transition. One can see a minor dip along $x$-direction in electric field distribution (Figs.~\ref{fields}d and~\ref{fields}h). It corresponds to the convex form of the EFC in the vicinity of $k_y=0$. Indeed, the purely flat EFC leads to the narrow-focused unidirectional propagation of surface wave~\cite{high2015visible,yang2017hyperbolic}, similar to the self-collimation regime in photonic crystals~\cite{kosaka1999self}.

\begin{figure}[t]
  \centering
  \includegraphics[width=0.99\linewidth]{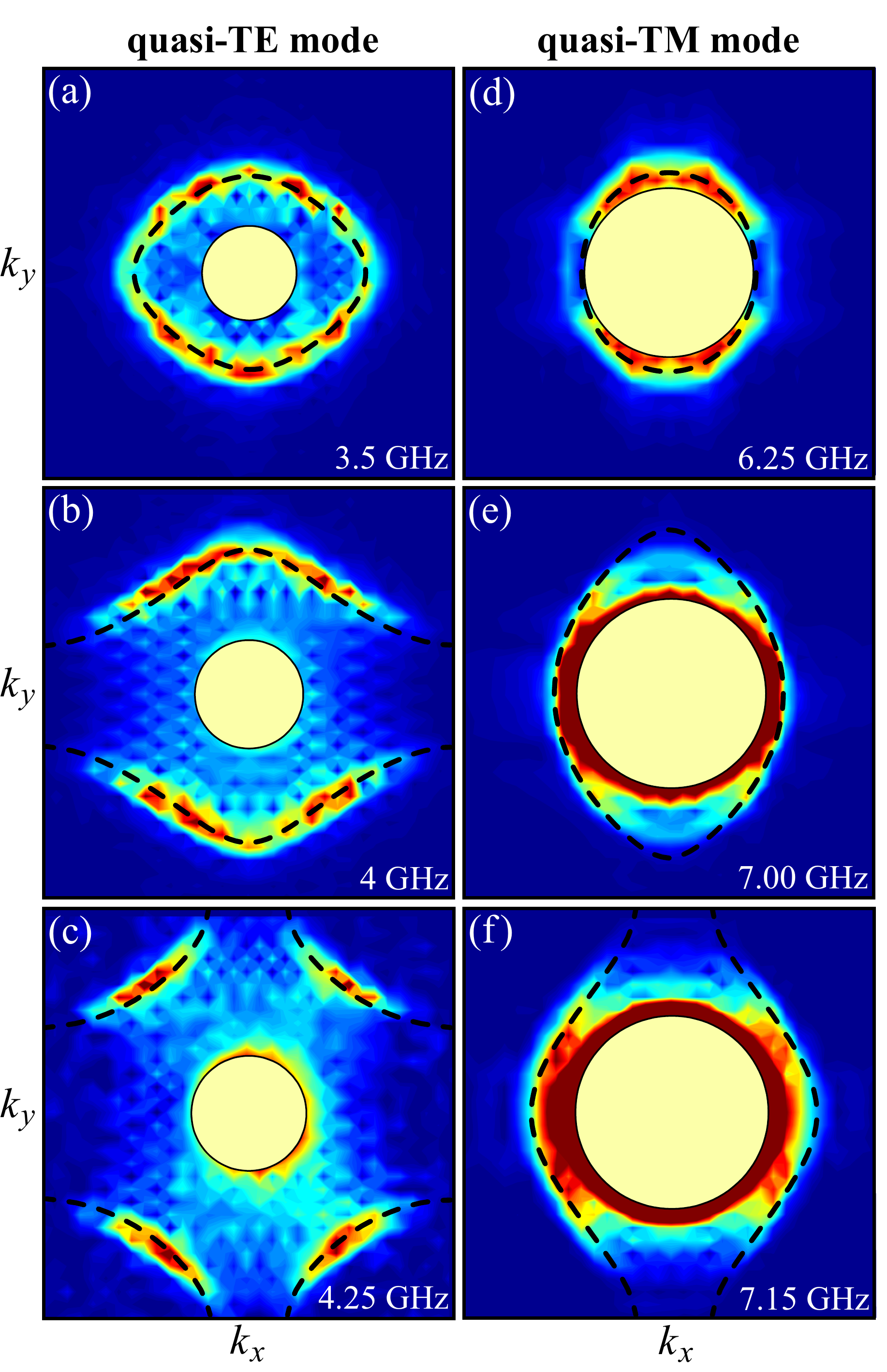}
  \caption{The equal frequency contours for the surface waves excited at the metasurface, depicted in Fig.~1, with the setup configuration 'loop-loop' (a-c) and 'probe-probe' (d-f) for different frequencies. Color maps correspond to the experimental measurements, dashed lines correspond to the numerical simulations with Eigenmode Solver in CST Microwave Studio. The frame is limited by the first Brillouin zone. The circles correspond to the light line.}
  \label{iso}
\end{figure}

\begin{figure*}[t]\centering
  \includegraphics[width=0.85\linewidth]{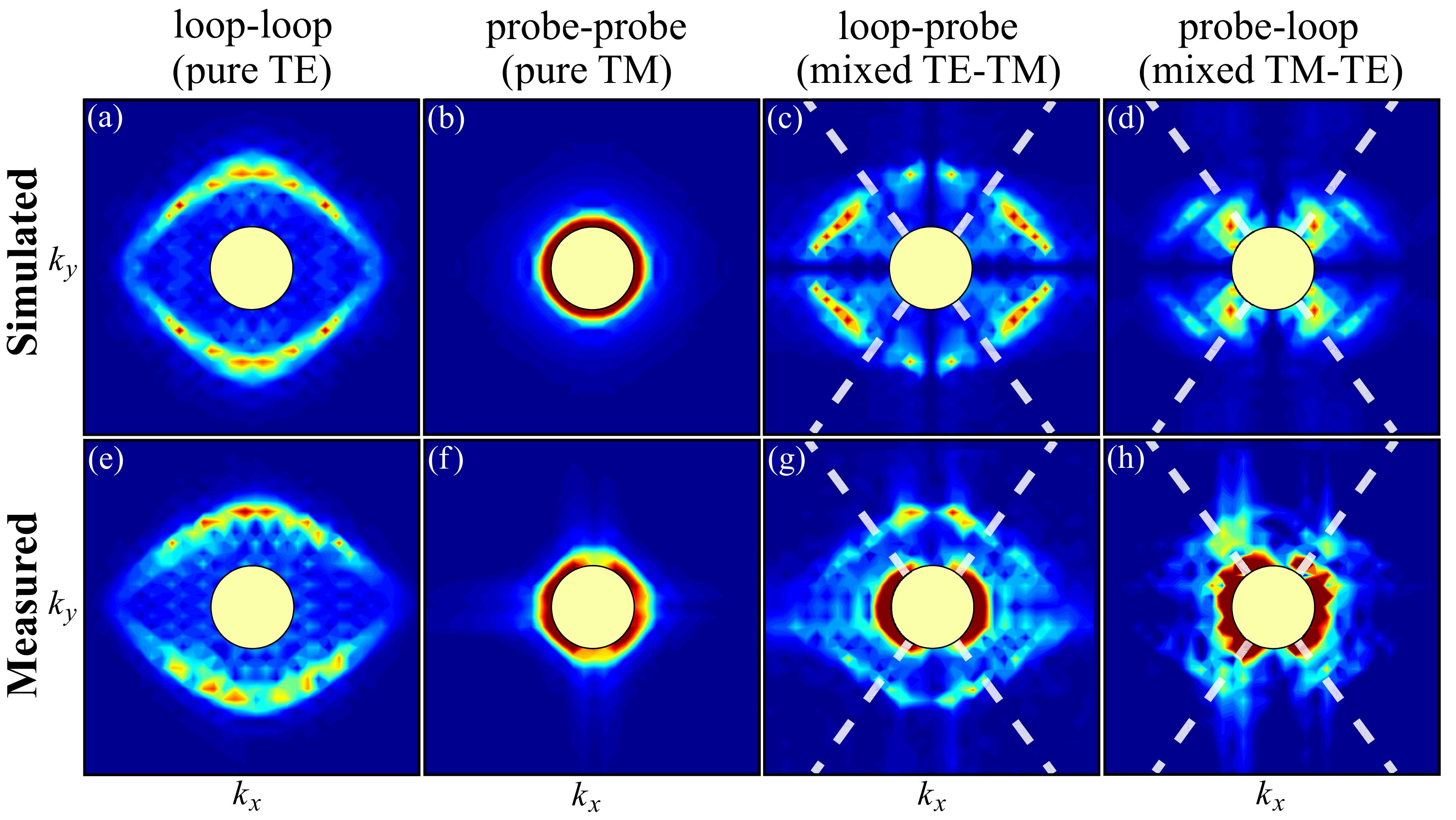}
  \caption{The equal frequency contours for the surface waves excited at the metasurface, depicted in Fig.~1, with different setup configurations at the frequency $3.75$ GHz, obtained numerically with CST Microwave Studio (a-d) and experimentally (e-h). The circles correspond to the light line. The dashed white lines correspond to the maximum hybridization direction.}
  \label{hybrid}
\end{figure*}

The experimental EFCs for the quasi-TE surface wave obtained via Fourier-transform of the magnetic field pattern measured with loop-loop setup configuration are shown in Figs.~\ref{iso}a-\ref{iso}c. The measured EFCs demonstrate a good correspondence with the simulated ones for all frequencies under consideration. One can follow up the evolution of the EFCs with the increase of the frequency, which is very similar to the one for the anisotropic plasmonic metasurface in the optical range~\cite{samusev2017polarization}. The quasi-TE mode can propagate at low frequencies and has the frequency limit according to the effective model~\cite{yermakov2015hybrid}. At the frequencies less than $2.5$~GHz the dispersion of the quasi-TE mode is close to the light line ($k_\tau < 1.5 \, \omega/c$) according to Fig.~\ref{fig:dispersion_cst}. At the frequency $f = 3.5$~GHz the EFC is closed and represents an ellipse (Fig.~\ref{iso}a), while it becomes opened at the frequency $f = 4$~GHz (Fig.~\ref{iso}b), i.e. topological transition takes place~\cite{krishnamoorthy2012topological}. It is related to the strong spatial dispersion and significantly changes from the EFCs predicted within effective medium approximation~\cite{yermakov2015hybrid} and the EFCs for the ultra-anisotropic case~\cite{yang2017hyperbolic}. It should be pointed out that surface wave dispersion depends on the propagation direction due to the anisotropy, which has the direct relation to the Dyakonov surface waves~\cite{dyakonov1988new,takayama2017midinfrared,takayama2018experimental}. The EFCs at the frequency of 4.25~GHz in Fig.~\ref{iso}c demonstrate the highly-directional surface waves excitation that can improve the collection efficiency of the emitters disposed in the vicinity of a metasurface in the way discussed in Ref.~[\onlinecite{kildishev2013planar}].

The EFCs for the quasi-TM surface wave obtained via Fourier-transform of the electric field pattern measured with the probe-probe setup configuration are shown in Figs.~\ref{iso}d-\ref{iso}f. One can see a good coincidence between numerical and experimental results. At low frequencies quasi-TM mode is very close to the light line that is in agreement with Fig.~\ref{fig:dispersion_cst}. The EFC at the frequency $f = 6.25$~GHz represents an ellipse, which is elongated along $y$-direction  and is very close to the light line along $x$-direction (Fig.~\ref{iso}d). At higher frequency this ellipse increases (Fig.~\ref{iso}e), then it becomes open and almost flat at 7.15~GHz (Fig.~\ref{iso}f). Finally, the EFC sharply transforms to the highly-directional near-corner arc at 7.25~GHz (Fig.~\ref{iso_cst}b).


In order to demonstrate the hybridization we compare the EFCs at the frequency $f=3.75$~GHz in different setup configurations for the numerical and experimental field patterns, which is shown in Fig.~\ref{hybrid}. The hybridization is explicitly proved in the mixed loop-probe and probe-loop setup configurations, where we can observe the measured normal magnetic field component from vertical electric dipole excitation (Figs.~\ref{hybrid}c and \ref{hybrid}g) and vice versa (Figs.~\ref{hybrid}d and \ref{hybrid}h). The surface waves of purely TE or TM polarizations (with no cross-polarization) can propagate only along the principle axes (along $x$~and~$y$), when the coupling factor of dispersion equation vanishes~\cite{yermakov2015hybrid}. For all other propagation directions the surface waves have a hybrid polarization due to the metasurface anisotropy, i.e. the electromagnetic field is a combination of TE and TM polarized contributions. So, we cannot observe the surface waves propagation along the principal axes for the loop-probe and probe-loop setup configurations that is confirmed by simulation in Figs.~\ref{hybrid}c and \ref{hybrid}d. The measured EFCs, shown in Figs.~\ref{hybrid}g and \ref{hybrid}h, also confirm reduction of the mixed wave field component along the principal axes. The incomplete suppression of these components in the experiment can be explained by some asymmetry of the used sources and probes. It is important to note that both hybrid TE-TM eigenmodes can propagate simultaneously at low frequencies, which is shown in Fig.~\ref{fig:dispersion_cst}. Despite the hybridization of pure TE and TM modes, only one type of polarization is predominant for each mode, frequency and propagation direction. One can see the propagation direction of the maximum hybridization degree, where TE polarization is mostly suppressed by the TM one. It corresponds to the dips of the field intensity for the loop-probe configuration (Figs.~\ref{hybrid}c and \ref{hybrid}g) and the appropriate peaks for the probe-loop configuration (Figs.~\ref{hybrid}d and \ref{hybrid}h).

\section{Conclusions}
In this work, we have experimentally studied the surface waves localized at a hyperbolic metasurface. We have analyzed the evolution of the equal frequency contours for both modes. We have observed different shapes of equal frequency contours, which correspond to a plethora of excitation and propagation regimes of surface waves. The drastic change of the equal frequency contour shape from closed to open, called topological transition, is observed. For the first time, we have experimentally demonstrated the existence of two modes with hybrid TE-TM polarization and their simultaneous propagation. We believe that the observed effects can be important for the efficient manipulation and control over electromagnetic surface waves, which is promising for a number of potential applications in optical information technologies, opto-electronic and photonic devices, on-chip networks, opto-mechanics and biological sensors.

\begin{acknowledgments}
This work was supported by Russian Foundation for Basic Research (N\textsuperscript{\underline{o}}16-37-60064, 17-02-00538, 17-02-01234, 18-32-00739), the Ministry of Education and Science of the Russian Federation (3.1668.2017/4.6, 3.8891.2017/8.9), Goverment of Russian Federation (Grant 08-08) and the Foundation for the Advancement of Theoretical Physics and Mathematics "BASIS". 
\end{acknowledgments}

\bibliography{bibliography}

\end{document}